\documentclass[usenatbib]{mn2e}

\usepackage{verbatim}
\usepackage{amssymb}
\usepackage{amsmath}
\usepackage{graphicx}
\usepackage[normalem]{ulem}

\usepackage[T1]{fontenc}  
\usepackage{aecompl}

\title[Magnetic fields in protoplanetary gaps]{Magnetic fields in gaps surrounding giant protoplanets}
\author[Sarah L. Keith and Mark Wardle]{Sarah L. Keith $^{1,2}$\thanks{E-mail: sarah.keith3@students.mq.edu.au; mark.wardle@mq.edu.au} and Mark Wardle$^{1}$\\$^{1}$Department of Physics \& Astronomy and Research Centre in Astronomy, Astrophysics \& Astrophotonics, Macquarie University,  \\NSW 2109, Australia\\$^{2}$Jodrell Bank Centre for Astrophysics, School of Physics and Astronomy, The University of Manchester, Manchester, M13 9PL,\\United Kingdom}
\begin{document}
\date{Accepted 2015 May 5.  Received 2015 May 4; in original form 2015 February 4}
\pagerange{\pageref{firstpage}--\pageref{lastpage}} \pubyear{Year}
\maketitle
\label{firstpage}\index{}

\begin{abstract}
Giant protoplanets evacuate a gap in their host protoplanetary disc, which gas must cross before it can be  accreted. A magnetic field is likely carried into the gap, potentially influencing the flow. Gap crossing has been simulated with varying degrees of attention to field evolution (pure hydrodynamical, ideal, and resistive MHD), but as yet there has been no detailed assessment of the role of the field accounting for all three key non-ideal MHD effects: Ohmic resistivity, ambipolar diffusion, and Hall drift. 

We present a detailed investigation of gap magnetic field structure as determined by non-ideal effects. We assess susceptibility to turbulence induced by the magnetorotational instability, and angular momentum loss from large-scale fields.  

As full non-ideal simulations are computationally expensive, we take an a posteriori  approach, estimating  MHD quantities from the  pure hydrodynamical gap crossing simulation by \citet{2012ApJ...747...47T}. We calculate the ionisation fraction and estimate field strength and geometry to determine the strength of non-ideal effects.

We find that the protoplanetary disc field would be easily drawn into the gap and  circumplanetary disc. Hall drift dominates, so that much of the gap is conditionally MRI unstable depending on the alignment of the field and disc rotation axes. Field alignment also influences the strong toroidal field component permeating the gap.  Large-scale magnetic forces are small in the circumplanetary disc, indicating they cannot drive accretion there. However, turbulence will be key during satellite growth as it affects critical disc features, such as the location of the ice line.

\end{abstract}

\section{Introduction}
\label{sec:introduction}

Giant planets capture their massive  atmospheres from the surrounding  protoplanetary disc. As the planet grows, its gravitational sphere of influence expands so that gas is captured faster than the nebula can resupply it, and a  gap is evacuated in the nebula  around the planet \citep{1985prpl.conf..981L, 1996ApJ...467L..77A, 1999ApJ...514..344B}.   Gaps have been seen in $\sim\mu$m scattered light  around TW Hya and infrared polarimetry of surface dust around HD97048 \citep{2013ApJ...771...45D, 2014A&A...568A..40G}. Spiral waves, likely launched by a protoplanet, have also been seen in HCO$^+$ line emission within a gap in the HD 142527 circumstellar disc \citep{2013Natur.493..191C}. 

A circumplanetary disc encircles the planet following the collapse and detachment of the protoplanet envelope from the nebula \citep{1982Icar...52...14L, 2009MNRAS.397..657A}. Gas flowing across the gap supplies the circumplanetary disc and controls the protoplanetary accretion rate \citep{2006ApJ...641..526L}. Gas with too much angular momentum to reach the planet directly is captured by the circumplanetary disc and its flow is controlled by angular momentum transport processes operating within the circumplanetary disc (\citealp{2011ApJ...743...53F, 2012ApJ...749L..37L, 2012A&A...548A.116R, 2012ApJ...747...47T}, hereafter TOM12; \citealp{2014ApJ...785..101F, KeithWardle2014}, hereafter KW14; \citealp{2014ApJ...782...65S, 2014ApJ...783...14T}). The circumplanetary flow pattern potentially comprises of high-altitude inflow, and midplane Keplerian and outflow components (\citealp{2003MNRAS.341..213B, 2003ApJ...586..540D}, TOM12, \citealp{2013ApJ...779...59G}). Although circumplanetary discs are yet to be observed, there are prospects for their detection through infrared spectral energy distributions observed with the James Webb Space Telescope or the Atacama Large Millimeter/submillimeter Array (ALMA; \citealp{2014ApJ...788..129I, 2015MNRAS.448L..67D, 2015ApJ...799...16Z}). This is strengthened by the recent observation of discs around three brown dwarfs, in  infrared continuum and CO line emission \citep{2014ApJ...791...20R}.

Magnetic fields likely play a role in the dynamics of the gap and circumplanetary disc system (\citealp{2013ApJ...779...59G, 2013ApJ...769...97U}, KW14, \citealp{2014ApJ...783...14T}). The field can have numerous effects, including generating hydromagnetic turbulence via the magnetorotational instability (MRI; \citealt{1991ApJ...376..214B, 1995ApJ...440..742H}), centrifugally driven disc winds and jets \citep{1982MNRAS.199..883B, 1993ApJ...410..218W}, and magnetic braking \citep{2004ApJ...616..266M}. 

Magnetic forces are transmitted by charged particles, and so these mechanisms require a minimum ionisation fraction. Collisions between charged and neutral particles give rise to non-ideal effects which counteract flux freezing - Ohmic resistivity, Hall drift, and ambipolar diffusion. Sufficiently strong resistivity and ambipolar diffusion decouple the motion of the gas and field (see \citealp{2014prpl.conf..411T} and references within), and Hall drift can lead to complex field evolution sensitive to the global field orientation (\citealp{1999MNRAS.307..849W, 2012MNRAS.422.2737W}; \citealp{2013MNRAS.434.2295K}; \citealp{2014A&A...566A..56L}, \citealp{2014MNRAS.441..571O}, \citealp{2015ApJ...798...84B}).

Gap-crossing dynamics is complex and so studies have relied on numerical simulations to model the flow. These include both grid-based (\citealp{1999ApJ...514..344B, 1999ApJ...526.1001L, 2003MNRAS.341..213B, 2006ApJ...641..526L}, TOM12) and smoothed particle \citep{1999ApJ...514..344B, 2009MNRAS.393...49A, 2012MNRAS.427.2597A}  hydrodynamical simulations, and ideal magnetohydrodynamical (MHD) simulations \citep{2003MNRAS.339..993N, 2004MNRAS.350..829P, 2013ApJ...769...97U}. Global 3D MHD simulations of the wider system have also been used to generate synthetic observation maps  for ALMA \citep{2015A&A...574A..68F}.

The recent inclusion of Ohmic resistivity adds an important level of realism to gap-crossing modelling \citep{2013ApJ...779...59G}. Ohmic resistivity is strong in the dense circumplanetary disc, and can produce an MRI stable, dead zone with implications for planet growth and disc evolution. However, non-ideal effects are also at work in relatively low density regions where the Hall effect and ambipolar diffusion can be strong \citep{2007Ap&SS.311...35W}. Their inclusion may enhance or suppress turbulence, affecting the flow and protoplanet growth rate.Indeed, the addition of ambipolar diffusion  to resistive simulations radically alters the protoplanetary disc flow between 1--5\,au, producing laminar accretion powered by a magnetocentrifugal wind, rather than the traditional accreting turbulent surface layers \citep{2015ApJ...801...84G}.

In this article we perform a comprehensive study of the relative importance of the three non-ideal effects on fluid dynamics throughout the gap. Owing to the significant computational cost of including all three effects in a full, 3D simulation, we take an a posteriori, semi-analytic approach. We calculate MHD quantities using a snapshot from the TOM12 3D hydrodynamical simulation (described in Section \ref{subsec:disc_model}). This allows us to calculate detailed ionisation maps including ionisation from cosmic-rays, stellar X-rays and radioactive decay,  accounting for grain charging (Sections \ref{subsec:ionisation} and \ref{subset:ionisation_results}). We determine the strength of non-ideal effects (Sections \ref{subsec:coupling} and \ref{sec:coupling_ls_results}) to estimate the field strength and geometry (Sections \ref{subsec:magnetic_field} and \ref{subsec:magnetic_field_results}). We find that the field would not alter the general fluid motion, justifying our use of an underlying hydrodynamical gap-crossing model (i.e., TOM12).  The implications of incorporating magnetic fields and non-ideal effects in gap crossing, such as the additional force from large-scale fields,  are considered  (Sections \ref{subsec:field_tangling_length}  and \ref{subsubsec:magnetic_forces}). Finally, we present a summary and discussion of findings in Section \ref{sec:discussion}.

\section{Model description}
\label{sec:method}
In this section we outline the disc model used to describe the protoplanetary disc and gap. We give details of the ionisation and diffusivity calculations, along with estimates for the magnetic field strength and geometry.

\subsection{Disc and gap structure}
\label{subsec:disc_model}

We take a semi-analytic, a posteriori approach to the calculation. We use a snapshot of a pure hydrodynamical simulation as the basis for estimating  MHD quantities. 
The protoplanetary disc and gap  model that we use is the TOM12 three-dimensional, hydrodynamical simulation. The simulation models a protoplanetary disc surrounding a star of mass $M_*$, in which a gap has been carved out through  gas capture by an embedded protoplanet of mass $M_p$, at fixed orbital radius $d_p$. The simulation corresponds to a protoplanet with the present-day mass and orbital radius of Jupiter (i.e., $M_p=M_J$, and $d_p=d_J$, where $d_J\equiv5.2\,$au). Gas is treated as inviscid, and self-gravitational and magnetic forces are neglected. 

The planet is located at the origin of a local cartesian co-ordinate system $(x,y,z)$,  which orbits the star at an orbital angular frequency $\Omega_p$. The $x$-axis is oriented along the radial direction, $\hat{r}$, which extends from the star to the planet; the $y$-axis is oriented in the azimuthal direction, $\hat{\phi}$,  parallel to the the protoplanet orbit; and the $z$-axis is parallel to the protoplanetary disc angular momentum vector. To capture details of the fine flow structure near the protoplanet the simulation has eleven levels of nested grids, each with $n_x\times n_y \times n_z=64\times64\times16$ data points. The total simulated portion of the disc extends over  $x\in[-12H_p, 12H_p]$, $y\in[-12H_p/d_p, 12H_p/d_p]$, and $z\in [0, 6H_p]$ where $H_p$ is the scale height of the protoplanetary disc. The local cartesian approximation requires that the Hill radius, 
\begin{eqnarray}
R_H&=&d\left(\frac{M}{3 M_*}\right)^{\frac{1}{3}}\nonumber\\
&\approx&0.36\,\text{au\,} \left(\frac{d}{5.2\,\text{au}}\right) \left(\frac{M}{M_J}\right)^{\frac{1}{3}}\left(\frac{M_*}{M_\odot}\right)^{-\frac{1}{3}},
\end{eqnarray}
 which is roughly the feeding zone of the protoplanet, satisfies $R_H\ll d_p$. Here $M_\odot$ is the mass of the Sun. The simulation data used here was taken after 160.7  protoplanet orbits, allowing the simulation to approach a steady state. The gap reached a maximum density contrast relative to the unperturbed disc of $\sim3$.

The simulation used a non-dimensionalised form for the MHD equations and so we must rescale the velocity $\mathbf{v}=(v_x, v_y, v_z)$ and density $\rho$ data for our calculations. 
We rescale the column density using the minimum mass Solar nebula column density at the orbital radius of Jupiter $\Sigma_0=140\,\text{\,g\,cm}^{-2}$ \citep{2002ApJ...581..357K, 2011ARA&A..49...67W}. This model estimates the gas mass in the Solar nebula by augmenting the solid material in Solar System planets with sufficient hydrogen and helium to bring the composition up to Solar \citep{1977ApSS..51..153W, 1981PThPS..70...35H}.  
Observations of Class II Young Stellar Objects (i.e., those with discs and strong UV and H$\alpha$ emission indicating active accretion) are currently limited to $\sim 20\,$au resolution, however inferred disc column density profiles are broadly consistent with this profile. As the gap gas density naturally drops over time, we allow for simple reduction of the column density, $\Sigma(x,y,z)$, though multiplication of a constant parameter $f_{\Sigma}$, where $f_{\Sigma}=1$ recovers the original TOM12 results. 

The simulation is isothermal, and we adopt the temperature of a black body in radiative equilibrium with Solar luminosity, excluding any heating from inflow processes, at the orbital radius of Jupiter, $T=120\text{\,K}$ \citep{2007Ap&SS.311...35W}.  This is supported by interferometric CO images of T Tauri  and dust temperatures from SED modelling \citep{1998A&A...339..467G, 2005ApJ...631.1134A, 2007A&A...467..163P}.

We adopt Solar gas composition, with a mixture of 80\% molecular hydrogen and 20\% atomic helium. This corresponds to a mean molecular weight of $m_n=2.34m_p$, where $m_p$ is the mass of a proton. This allows us to calculate the neutral number density, $n$, isothermal sound speed  $c_s=0.65\,\text{km\,s}^{-1}$, and aspect ratio: 
\begin{equation}
\frac{H_p}{d}=\frac{c_s}{v_k}\text{ }=\text{ }4.5\times10^{-2}\left(\frac{d}{d_J}\right)^{\frac{1}{2}},
\end{equation}
where  $v_k=\left(GM_*/d\right)^{\frac{1}{2}}$ is the Keplerian velocity. 

The flow pattern within a gap transitions from orbiting the star [orbital radius $\sqrt{(x+d_p)^2+y^2}$], to orbiting the planet (orbital radius $\sqrt{x^2+y^2}$) within the Hill sphere. In our analytic calculations we approximate the flow geometry by a composite Keplerian angular velocity functions:
\begin{eqnarray}
\label{eq:composite_keplerian}
\Omega=\left\{
\begin{array}{lr}
\left[GM_p\left(x^2+y^2\right)^{-\frac{3}{2}}\right]^{\frac{1}{2}} & \text{for $\sqrt{x^2+y^2}\le R_H$,}\\
\left[GM_*(x+d_p)^{-3}\right]^{\frac{1}{2}} & \text{ otherwise}.
\end{array}
\right.
\end{eqnarray}

TOM12 also simulated the circumplanetary disc, however they note that as it is shear-dominated, artificial viscosity is strong and so the results are more reliable in the gap. Similarly we include the circumplanetary disc in our calculations with the caveat that the disc structure in this region is uncertain. 

The scale height in the circumplanetary disc,
\begin{eqnarray}
H_c&=&H_p\left(\frac{\sqrt{x^2+y^2 }}{3R_H}\right)^{\frac{3}{2}}\\
&\approx&5.2\times10^{-2}H_p\left(\frac{\sqrt{x^2+y^2}}{0.2\,R_H}\right)^{\frac{3}{2}},
\end{eqnarray}
is significantly lower, and this needs to be accounted for when calculating gradients in the fluid flow (see Section \ref{subsec:coupling}). To that end we use a composite scale-height, $H$, which transitions at the boundary of the circumplanetary disc hydrostatic region:
\begin{eqnarray}
\label{eq:composite_h}
H=\left\{
\begin{array}{llr}
H_c & \text{for $\sqrt{x^2+y^2}<0.2R_H$,  $z/H_c<5$, and}\\
H_p & \text{ otherwise}.
\end{array}
\right.
\end{eqnarray}

\subsection{Degree of ionisation}
\label{subsec:ionisation} 
Charged particles transmit magnetic forces to the bulk neutral flow. In the gap, ionisation is largely caused by penetrating cosmic-rays and stellar X-rays, however there is also a weak, pervasive contribution from decaying nuclides. 
We calculate the ionisation level by  solving the coupled rate equations for  electron and ion number densities and mean grain charge subject to overall charge neutrality. We follow the method outlined in KW14, but using a more detailed calculation of the grain charge. A summary of the method is given below. 

We calculate the grain number density, $n_g$, for spherical grains with radius $a_g=0.1\,\mu$m and bulk density 3\,g\,cm$^{-3}$  with a constant dust to gas mass ratio $f_{dg}\sim10^{-4}$ to account for incorporation of solids into protoplanetary bodies \citep{1994ApJ...421..615P}. 

Internal ionisation from radioactive decay is primarily from the short-lived radionuclide $^{26}$\!Al, which contributes an ionisation rate $\zeta_{\textrm{R}}=7.6\times10^{-19}$\,s$^{-1}\left(f_{dg}/f_{\odot}\right)$ \citep{2009ApJ...690...69U}. The ionisation rate has been scaled by the ratio of the dust mass fraction to the Solar photospheric metallicity, $f_{\odot}=1.3\%$, so that the $^{26}$\!Al abundance is consistent \citep{2009ARA&A..47..481A}. The midplane dust fraction will be enhanced by settling,  however the consequent increase in the ionisation fraction will be small as ionisation is predominantly by external radiation. 

External ionisation sources are attenuated  by the column density above and below  each location in the simulation. Cosmic-ray ionisation occurs at the interstellar cosmic-ray ionisation rate, $\zeta_{\textrm{CR}}=10^{-17}$\,s$^{-1}$, attenuated by the attenuation depth $\Sigma_{\textrm{CR}}=96$\,g\,cm$^{-2}$ approaching from above and below the disc \citep{1981PASJ...33..617U, 2009ApJ...690...69U}. We account for ionisation from diffuse scattered stellar X-rays but neglect direct illumination, instead assuming that shielding by the inner portion of the disc is effective. Ionisation for solar luminosity  stellar X-rays occurs at a rate $\zeta_{\textrm{XR}}=9.6\times10^{-17}$\,s$^{-1}\left(d/d_J\right)^{-2}$, with attenuation depth $\Sigma_{\textrm{XR}}=8$\,g\,cm$^{-2}$ \citep{1999ApJ...518..848I, 2008ApJ...679L.131T}. The total ionisation rate, $\zeta$, is the sum from the three sources $\zeta=\zeta_{\textrm{R}}+\zeta_{\textrm{CR}}+\zeta_{\textrm{XR}}$.

We solve for the electron, ion and grain charge number densities by calculating the grain charge needed for overall charge neutrality in the steady state [see equations (26)--(29) in KW14]. We include charge focussing in the electron and ion capture rate coefficients for grains \citep{1980PASJ...32..405U}:
\begin{eqnarray}
k_{ig}&=&\pi a_g^2 \sqrt{\frac{8k_b T}{\pi m_i}}\times\left\{\begin{array}{lr}
\exp\left({-\frac{q^2|Z_g|}{a_gk_bT}}\right)& Z_g<0\\
\left(1+\frac{q^2|Z_g|}{a_gk_bT}\right)& Z_g>0
\end{array}
\right.
\label{eq:kmg}\\
k_{eg}&=&\pi a_g^2 \sqrt{\frac{8k_b T}{\pi m_e}}\times\left\{\begin{array}{lr}
\left(1+\frac{q^2|Z_g|}{a_gk_bT}\right)&Z_g<0\\
\exp\left({-\frac{q^2|Z_g|}{a_gk_bT}}\right)& Z_g>0
\end{array}
\right.
\label{eq:keg}
\end{eqnarray}
where $Z_gq$ is the mean grain charge, $k_b$ is Boltzmann's constant, and $m_e$ is the electron mass. We model metals as a single species, adopting the mass, $m_i$, and abundance, $x_i$, of the most abundant metal - magnesium \citep{CRC, 2009ARA&A..47..481A}. The recombination rate for ions and electrons is $k_{ei}=1.20\times10^{-12}\,\left(T/1000\,\text{K}\right)^{-0.7}\text{cm}^3\,\text{s}^{-1}$ \citep{2013A&A...550A..36M}.
If the grain charge is known these equations give the ion and electron number densities:
\begin{eqnarray}
n_e&=&\frac{n_g k_{ig}}{2 k_{ei}}\left[\left(1+\frac{4k_{ei}n\zeta}{k_{eg}k_{ig}n_g^2}\right)^{\frac{1}{2}}-1\right]\\
n_i&=&n_e\frac{k_{eg}}{k_{ig}}. \label{eq:ni}
\end{eqnarray}

Equations (\ref{eq:kmg})--(\ref{eq:ni}) are analogous to equations (30)--(33) in KW14 except for the inclusion of a non-zero grain charge for charge focussing. The protoplanetary disc has a higher average ionisation fraction resulting in a higher grain charge which cannot be neglected (as it could be in the KW14 circumplanetary disc analysis). We determine the equilibrium values of $n_e, n_i$, and $Z_g$ numerically by solving equations (\ref{eq:kmg})--(\ref{eq:ni}) along with charge neutrality, for the grain charge. We use the Brent-Dekker method, \texttt{gsl\_root\_fsolver\_brent}, for root-finding implemented in the GNU Scientific Library to solve for the grain charge to an accuracy of 1\%  \citep{GSL}. This method combines bisection and secant methods for rapid convergence. 

We also determine the effectiveness of collisional ionisation produced in MRI turbulent regions. Currents generated by an MRI field may be accelerated sufficiently to ionise neutral particles, providing additional ionisation \citep{2012ApJ...760...56M, 2015ApJ...800...47O}. We calculate the kinetic energy of accelerated particles using equations (37)--(38) in KW14 to determine if it exceeds the ionisation potential of neutral species (e.g., potassium has the lowest ionisation energy of the abundant elements,  with $E\sim4.34$\,eV). 
We find that although electrons reach energies of up to 1\,eV in the disc atmosphere ($z\sim5H_p$) where  collisions are infrequent and the mean-free path is long,  in the denser midplane electrons 
are not accelerated above $0.05$\,eV. This is too low to ionise any atomic species and so the MRI does not contribute significantly to the ionisation. Thus, we neglect the effect in our calculation of the ionisation fraction.

Finally, we note that thermal ionisation is likely not effective in the gap surrounding Jupiter. The protoplanetary disc at $d=d_J$ is much cooler than the required $T\sim1000$\,K, and further shock heating of accreting material at the circumplanetary disc is weak. In the circumplanetary shock, 
shock heating flux $F_{\text{s}}\sim\sigma(40\,\text{K})^4$ (calculated for an accretion rate $10^{-6}M_J/$year, at $r\sim 300R_J$ from the protoplanet; \citealp{1981Icar...48..353C})
is negligible compared with Solar irradiation, $F_{\odot}=\sigma(120\,\text{K})^4$  at this orbital radius, precluding thermal ionisation.

\subsection{Non-ideal MHD effects and magnetic coupling}
\label{subsec:coupling}
The evolution of the magnetic field and its role in the gas dynamics depends on the strength and nature of the magnetic coupling between the field and gas. Magnetic coupling is manifest in the induction equation,
\begin{eqnarray}
  \frac{\partial \bold{B}}{\partial t}&=&\nabla\times(\bold{v}\times \bold{B})-\nabla\times[\eta_O(\nabla\times \bold{B})+\eta_H(\nabla\times \bold{B})\times \hat{\bold{B}}]\nonumber\\
&&-\nabla\times[\eta_A(\nabla\times \bold{B})_{\perp}],
\label{eq:induction}
\end{eqnarray}
where $\eta_O$, $\eta_H$, and $\eta_A$ are the transport coefficients for Ohmic resistivity,  Hall drift, and ambipolar diffusion. These non-ideal effects cause the magnetic field to slip through the gas, and can dominate the field evolution if they are too strong. They are caused by neutral collisions disrupting the $\mathbf{E\times B}$ drift of charged particles. Differences in the density dependence of these effects leads to three different regimes in which Ohmic resistivity,  Hall drift and ambipolar diffusion are most effective in high, moderate, and low density regions, respectively. For a predominantly vertical magnetic field, Ohmic resistivity and Ambipolar diffusion act similarly (and distinctly from the Hall drift), and so they are often combined into the Pedersen diffusion $\eta_P=\eta_O+\eta_A$. We calculate the strength of the three effects using  the transport coefficients from equations (53)--(55) in KW14\footnote{There is a typographical error in the expression for Ohmic resistivity in equation (53) of KW14. The scaling of the diffusivity with the neutral density should be $\eta_O\propto \rho$ rather than the inverse proportionality given. Additionally, the grain Hall Parameter in equation (51) of KW14 is a factor of $\sqrt{m_g/m_n}$ too large, and should be written as $\beta_g=5.5\times10^{-8}Z_g\left(B/1\,\text{G}\right)\left(10^{15}\,\text{cm}^{-3}/n\right)\left(0.1\,\mu\text{m}/a_g\right)^{2}\sqrt{1000\,\text{K}/T}$, instead. This reduction has no impact on the KW14 results, as $\beta_g$ was already so small as to be negligible.}.

How strong these effects must be to dominate the inductive term and decouple the motion of the field and gas depends on the field geometry. Below we outline the coupling conditions for MRI turbulent, toroidal, and poloidal components in turn. 

The MRI dynamo harnesses shear to generate a turbulent field. Diffusion restricts MRI turbulence by limiting bending of the smallest-scale field modes.  The MRI operates effectively if the fastest growing mode, with wavelength $\lambda=2\pi v_a/\Omega$, survives diffusion. Magnetic diffusion counteracts dynamo growth of short wavelength fluctuations with wavelength $\lambda<\eta/v_{a,z}$, for which the diffusion rate exceeds the growth rate. Here $v_{a,z}=B_z/\sqrt{4\pi\rho}$ the vertical Alfv\'{e}n speed.  Thus,  the MRI operates in effectively ideal MHD conditions if Pedersen and Hall components are simultaneously below the threshold \citep{2002ApJ...577..534S, 2014prpl.conf..411T}:
\begin{equation}
\eta_P\lesssim v_{a,z}^2\Omega^{-1} \text{ and } \lvert\eta_H\rvert \lesssim v_{a,z}^2\Omega^{-1}.\label{eq:MRI_criterion1}
\end{equation}
Here, the absolute value of the Hall term reflects its signed nature. The MRI also requires the field must be weak enough that the fastest-growing mode is confined within the disc scale height, so that for ideal MRI \citep{2011ApJ...742...65O}:
\begin{equation}
\beta_z=\frac{8\pi\rho c_s^2}{B_z^2}>8\pi^2. \label{eq:betaz_constraint}
\end{equation}

Unlike Ohmic resistivity and ambipolar diffusion,  Hall drift is not diffusive\footnote{Indeed, \citet{2013MNRAS.434.2295K} found that the Hall effect can be \textit{anti-diffusive} in unstratified shearing boxes.} and can change the behaviour of MRI turbulence, even in the presence of strong Ohmic and ambipolar diffusion. Hall drift is antiparallel to the current density  and may cooperate with, or act against Keplerian shear according to the relative orientation of the protoplanetary disc angular momentum vector and  vertical component of the magnetic field, $s=\text{sign}(\bold{B}\cdot\bold{\Omega})$ \citep{1999MNRAS.307..849W, 2001ApJ...552..235B, 2012MNRAS.422.2737W}. For example, if the field is aligned with the rotation axis, the Hall effect can destabilise the flow, enhancing turbulence. On the other hand, if the field and rotation axis are anti-aligned, Hall drift acts against the shear, suppressing turbulence. In the Hall-MRI regime, the Hall effect  counteracts  Pedersen stabilisation against field tangling if it dominates and the vertical field orientation is favourable \citep{2012MNRAS.422.2737W, 2014A&A...566A..56L}. Hall-MRI  requires
\begin{equation}
\lvert\eta_H\rvert >\eta_P\text{, }\lvert\eta_H\rvert>v_{a,z}^2\Omega^{-1}\text{, and }s=1,\label{eq:MRI_criterion2}
\end{equation}
along with a Hall-MRI weak-field criterion, which we show in Appendix \ref{appendix:weak_field} is the same as  equation (\ref{eq:betaz_constraint}). Indeed, the Hall effect need not dominate diffusion to impact turbulence,  as even weak Hall drift modifies the wave number and growth rate of the fastest growing mode \citep{2012MNRAS.422.2737W}.

A toroidal field surrounding the star or protoplanet, must also be maintained against diffusion which would tend to unwrap and straighten field lines. The impact of non-ideal effects on a toroidal filed are most important close to the central object where  gradients are strongest. A toroidal field is preserved against diffusion if magnetic induction exceeds  non-ideal effects, as captured by the coupling threshold (c.f., \citealp{2008ApJ...679L.131T}; see Appendix \ref{appendix:toroidal_hall}):
\begin{equation}
\eta_P\lesssim  \Omega H^2\text{, and }\lvert\eta_H\rvert \lesssim  \Omega H^2.\label{eq:ideal_toroidal_criterion1}
\end{equation} 

A toroidal field can also be preserved against strong diffusion if  Hall drift dominates. Hall drift is along the current, which is in the fluid rotation direction for both the aligned and anti-aligned  cases (i.e., $s=1$ and $s=-1$).  So, either by enhancing the field wrapping (i.e., for $s=1$) or by unwinding and rewrapping the field in the counter flow direction (i.e., for $s=-1$), the Hall effect produces a strong negative toroidal field.  See Appendix \ref{appendix:toroidal_hall} for a discussion of the effect of vertical field direction dependence for coupling of a toroidal field.  Simulations are needed to verify this behaviour, but we highlight Hall-dominated regions with the anticipation it may extend the reach of the toroidal field. We simply require that azimuthal winding from  Hall drift exceeds diffusion:
\begin{equation}
\lvert\eta_H\rvert >\eta_P\text{, }\lvert\eta_H\rvert >\Omega H^2\label{eq:ideal_toroidal_criterion2}.
\end{equation} 

Finally, the disc contains a poloidal field component throughout. Although magnetic coupling limits the minimum field bending radius  of the poloidal component, an essentially vertical field will always couple, owing to its small gradient. As we are not concerned with the exact geometry of the poloidal component, we simply assume that it is approximately vertical, couples everywhere and so permeates the gap. 

\subsection{Magnetic field strength and geometry}
\label{subsec:magnetic_field}
Protoplanetary discs inherit a large-scale magnetic field from their progenitor molecular cloud. Field measurements are difficult as current observations are limited to  $\sim 100\,$au resolution \citep{2014Natur.514..597S}. The disc field is certainly compressionally enhanced over the cloud field ($B\sim1$--$100$\,mG; \citealp{1987ARA&A..25...23S}),  while the equipartition field $B_{\text{eq}}=8\pi p$ provides a maximum field strength the disc will support before magnetic forces exceed the thermal pressure, $p=\rho c_s^2$.

Nevertheless,  it is possible to gain a more precise estimate from the stellar accretion rate, $\dot{M}\sim10^{-8}\,M_{\odot}$\,yr$^{-1}$, as the field is thought to play a principle role in angular momentum transport (\citealp{1998ApJ...495..385H}; see references in \citealp{2007ARA&A..45..565M}). This can be made from the azimuthal component of the axisymmetric momentum equation, vertically integrated between the disc surfaces \citep{2007Ap&SS.311...35W},
\begin{equation}
\rho\left[\left(\mathbf{v}\cdot\mathbf{\nabla}\right)\mathbf{v}\right]_{\phi}=\frac{\left(\mathbf{B}\cdot\mathbf{\nabla}\mathbf{B}\right)_{\phi}}{4\pi}.
\end{equation}
This yields a  minimum field strength for  vertical, toroidal, and turbulent components in the disc  \citep{2007Ap&SS.311...35W}:
\begin{eqnarray}
B_{z}&=&6.0\times10^{-3}\,\text{G}\left(\frac{\dot{M}}{10^{-8}\,M_{\odot}\,\text{yr}^{-1}}\right)^{\frac{1}{2}}\nonumber\\
&&\times\left(\frac{\Omega}{1.7\times10^{-8}\text{\,s}^{-1}}\right)^{\frac{1}{2}}\left(\frac{d}{d_p}\right)^{-\frac{1}{2}},\label{eq:bz_disc}\\
B_{\phi}&=&2.8\times10^{-2}\text{\,G}\left(\frac{\dot{M}}{10^{-8}\,M_{\odot}\,\text{yr}^{-1}}\right)^{\frac{1}{2}}\nonumber\\
&&\times\left(\frac{\Omega}{1.7\times10^{-8}\text{\,s}^{-1}}\right)^{\frac{1}{2}}\left(\frac{H_p}{0.23\text{\,au}}\right)^{-\frac{1}{2}}\text{, and} \label{eq:bphi_disc}\\
B_{\text{MRI}}&=&5.7\times10^{-2}\text{\,G}\left(\frac{\dot{M}}{10^{-8}\,M_{\odot}\,\text{yr}^{-1}}\right)^{\frac{1}{2}}\nonumber\\
&&\times\left(\frac{\Omega}{1.7\times10^{-8}\text{\,s}^{-1}}\right)^{\frac{1}{2}}\left(\frac{H_p}{0.23\text{\,au}}\right)^{-\frac{1}{2}}\label{eq:bmri_disc} 
\end{eqnarray}
respectively.  Components of the turbulent field are \citep{2004ApJ...605..321S}:
\begin{eqnarray}
B_{\text{MRI},r}&=&0.35B_{\text{MRI}},\label{eq:projected_MRIr}\\
B_{\text{MRI},\phi}&=&0.92B_{\text{MRI}}\text{, and }\label{eq:projected_MRIphi}\\
B_{\text{MRI}, z}&=&0.19 B_{\text{MRI}}.\label{eq:projected_MRIz}
\end{eqnarray}

We now turn to estimating the magnetic field strength in the gap. Turbulence would be continuously generated there so that equations (\ref{eq:bmri_disc})--(\ref{eq:projected_MRIz}) remain valid, however our estimates of the vertical and toroidal fields are not sufficient in the gap.

The vertical component would be drawn in with the flow and reduced or enhanced in keeping with the large density variation across the gap. We assume that the field is frozen into the gas at the outer edge of the gap (at $x_g=1.45$\,au), drawn into the gap by the flow.  The self-consistency of our adoption of this assumption is verified by our results (see Sections \ref{sec:coupling_ls_results}, \ref{subsubsec:magnetic_forces}). The field would be pinned to the flow at the lowest magnetically-coupled gas layer, reducing or enhancing the field according to $B\propto\Sigma^{-1}$.  Nevertheless, for simplicity, we assume the field is pinned at the midplane,  with a field strength $B_{z, g}=4.4\times10^{-3}$\,G, given by equation (\ref{eq:bz_disc}) at $x_g=1.45$\,au, $y=0$ where $\Sigma_g=97\text{\,g\,cm}^{-2}$. Normalising to the field strength at the outer edge of the gap, this leads to a vertical field strength profile of
\begin{equation}
B_{z}=B_{z, g}\left(\frac{\Sigma}{\Sigma_g}\right)\label{eq:fc_z}
\end{equation}
in the gap. 

Determining the toroidal field strength is considerably more difficult as the field would be wound up continually until field-line drift balances differential rotation. As this is beyond the scope of our treatment, we simply assume the constant ratio $\sqrt{d/H_p}$ between the toroidal and vertical components in the disc is preserved for our estimate in the gap. This leads to a gap toroidal field strength profile of:
\begin{equation}
B_{\phi}=B_{\phi, g}\left(\frac{\Sigma}{\Sigma_g}\right), \label{eq:fc_phi}
\end{equation}
where $B_{\phi, g}=2.0\times10^{-2}$\,G is the toroidal field strength, likewise taken at $x_g=1.45$\,au, $y=0$. We cap the field strength so that the associated magnetic pressure does not exceed the gas pressure:
 \begin{eqnarray}
\beta_{\phi}&=&\frac{8\pi\rho c_s^2}{B_{\phi}^2}\ge1.\label{eq:beta_phi}
\end{eqnarray}
This allows the field to expand in response to strong magnetic pressure at high altitude,  correspondingly reducing the field strength. 

Numerical simulations  suggest that the toroidal component may  be subject to periodic polarity reversals, as it is lost through magnetic buoyancy every $\sim40$ orbits \citep{2000ApJ...534..398M,2010ApJ...708.1716S,2011ApJ...732L..30H, 2012ApJ...744..144F}.
The magnitude of the toroidal field would also vary vertically between its surface value, $B_{\phi,s}$, and zero at the midplane, with the approximate scaling $B_{\phi}\sim B_{\phi,s}(z/H_p)$, appropriate for a thin, rotationally-supported disc that is well coupled to the magnetic field \citep{1993ApJ...410..218W}. 
  For simplicity we  take $B_{\phi}$ constant with height, but account for vertical gradients (e.g., in Appendix \ref{appendix:toroidal_hall}), and probe the effect of uncertainty by global enhancement or reduction of the the poloidal and toroidal components through multiplication by a constant parameter $f_B$. 

\subsubsection{Self-consistent coupled field geometry}
\label{sec:self_consistent_field}
The variation of diffusion and field gradients across the gap  means that not all three field components are present everywhere.  Here we determine where diffusion permits each field component. We use the diffusivities to determine the magnetic structure based on the level of coupling for each field component (poloidal, toroidal, and turbulent), and only include the components where they couple. Our procedure for calculating the field geometry is: 
\begin{enumerate}
\item Poloidal field - we include a poloidal component at all locations, calculated with equation (\ref{eq:bz_disc}) in the main protoplanetary disc (i.e., $\lvert x\rvert>x_g$),  or using the flux-conserved form, equation (\ref{eq:fc_z}), inside the gap. It is safe to include a poloidal component everywhere since a pure vertical field always couples and that any radial bending will adjust to the level of diffusion. 
\item Toroidal field - we calculate the toroidal component strength using equation (\ref{eq:bphi_disc}) in the bulk protoplanetary disc flow, and equation (\ref{eq:fc_phi}) in the gap. We cap the field strength by the gas pressure through equation (\ref{eq:beta_phi}). We calculate $\eta_O$, $\eta_H,$ and $\eta_A$ using the net field strength, $B=\left(B_z^2+B_{\phi}^2\right)^{1/2}$, and determine if the field is coupled using equations (\ref{eq:ideal_toroidal_criterion1}) and (\ref{eq:ideal_toroidal_criterion2}). If the field is coupled the toroidal component is kept; otherwise it is set to zero. 
\item MRI turbulent field - finally, we determine if an MRI turbulent field is sustained. We calculate the turbulent field strength in  both the protoplanetary disc and the gap using equations (\ref{eq:bmri_disc})--(\ref{eq:projected_MRIz}). We calculate $\eta_O$, $\eta_H,$ and $\eta_A$  using the net field strength, $B=\left[B_{\text{MRI}, r}^2 + (B_{\phi}+B_{\text{MRI}, \phi})^2+(B_z+B_{\text{MRI}, z})^2\right]^{1/2}$, and determine if the field is sustained using equations (\ref{eq:MRI_criterion1})--(\ref{eq:betaz_constraint}).
\end{enumerate}

\section{Results}
\label{sec:results}
Here we present the results of the calculations developed in Section \ref{sec:method}, including the ionisation fraction, magnetic field, and strength of non-ideal MHD effects. Figures are shown for a Jupiter-mass planet, orbiting a Solar mass star, at the present orbital distance of Jupiter. Unless otherwise stated, we take $f_{dg}=10^{-4}$, $f_{\Sigma}=1$,  $f_B=1$, and $s=1$.  

\subsection{Degree of ionisation}
\label{subset:ionisation_results}
\begin{figure*}
\begin{center}
\includegraphics[width=0.98\textwidth]{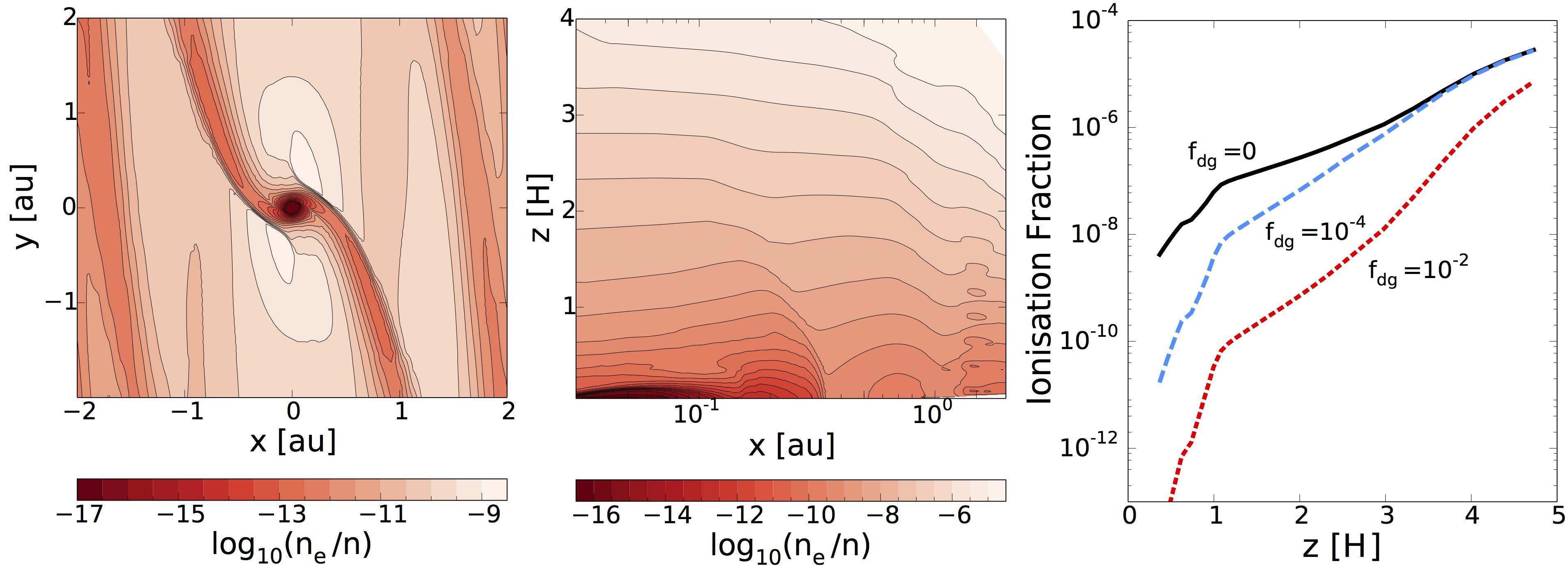}
\end{center}
\caption{\label{fig:ionisation} Ionisation fraction profile slices through the gap, a posteriori computed  for the hydrodynamical simulation of \citet{2012ApJ...747...47T}.  Left and centre panels show 2D slices of the ionisation fraction $\log_{10}(n_e/n)$ as contour plots with logarithmically spaced contours for $y=0$ and $z=0$, respectively, and for $f_{dg}=10^{-4}$.  Right panel shows $n_e/n$ as a function of height for $x=0.3\,$au$, y=0$ for dust to gas mass ratios $f_{dg}=0, 10^{-4}, 10^{-2}$ with solid, dashed, and dotted curves respectively. }
\end{figure*}

Fig. \ref{fig:ionisation} shows the ionisation profiles within the gap and disc. The left and centre panels show 2D slices through the gap, with the colour scale showing the ionisation fraction. The left panel shows the top-down view in the $x$--$y$ plane at the midplane, and the centre panel shows the edge-on profile in the $x$--$z$ plane at $y=0$. The star is located beyond the simulation boundary at $x=-d_J$, $y=0$. Small white patches in the top- and bottom-right corners of the edge-on profiles are beyond the simulation boundary.
The right panel shows the dependence of the vertical ionisation profile on the dust to gas mass ratio, at $x=0.3$\,au, $y=0$. 

Shielding from the overlying gas reduces the ionisation fraction with height and in denser regions. X-rays penetrate  to $z\sim H_p$, while cosmic rays reach everywhere except for the circumplanetary disc (located at $x, y \lesssim0.1$\,au), owing to its high column density. 
Instead, ionisation in the circumplanetary disc is from  radioactive decay, which is a much weaker ionising source. 
Consequently, the ionisation fraction in the circumplanetary disc is much weaker than in the protoplanetary disc, as supported by previous studies (\citealt{2011ApJ...743...53F, 2014ApJ...785..101F},  KW14, \citealt{2014ApJ...783...14T}).
The sharp drop in the ionisation fraction profile  at $z\sim H_p$ occurs at the shock boundary surrounding the circumplanetary disc, visible as a halo of poorly ionised gas in the edge-on view shown in the centre panel of Fig. \ref{fig:ionisation}. 

Dust grains reduce the ionisation fraction by soaking up free electrons. Grain charge capture is significant and most efficient at the midplane where grains are abundant. It also reduces radioactive decay, influencing ionisation, and hence the potential for magnetic coupling, in the circumplanetary disc.


\subsection{Non-ideal MHD effects and magnetic coupling}
\label{sec:coupling_ls_results}

\begin{figure*}
\begin{center}
\includegraphics[width=0.98\textwidth]{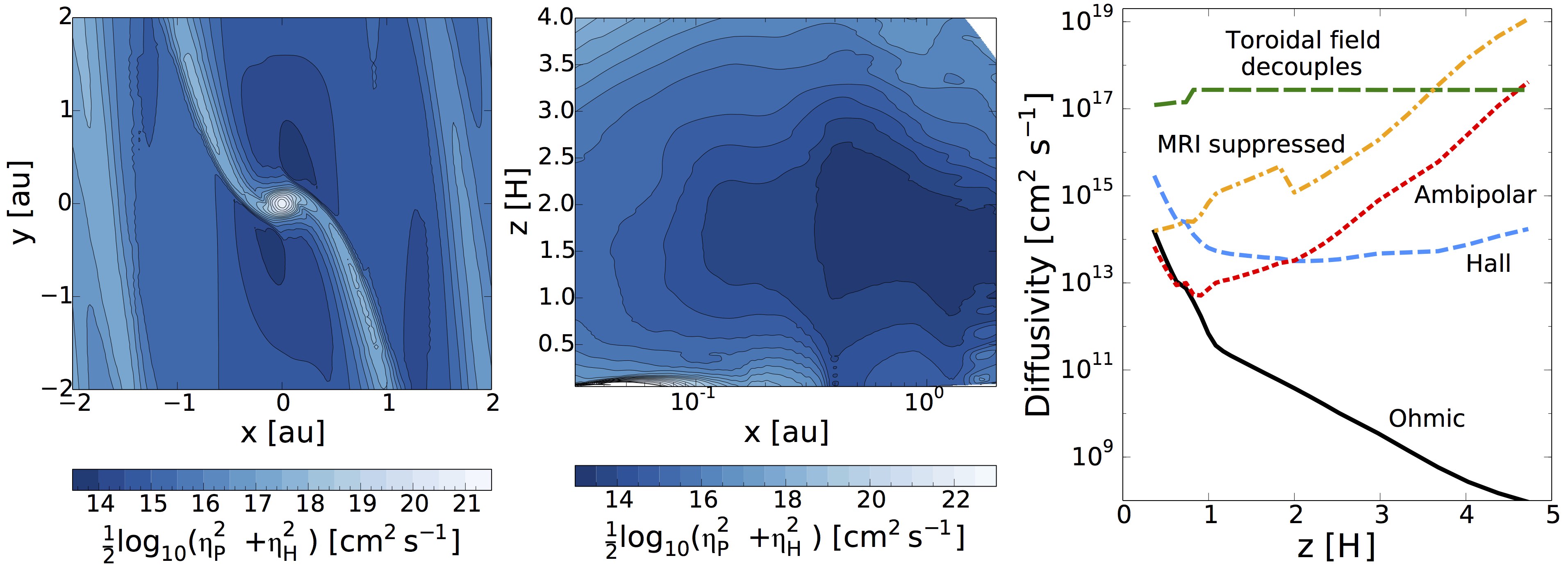}
\end{center}
\caption{\label{fig:diffusivity}  As for Figure \ref{fig:ionisation} but showing a contour plot of the field-perpendicular diffusivity, $\eta_{\perp}=\sqrt{\eta_P^2+\eta_H^2}$ at the midplane (left panel), and $y=0$ (centre panel). Ohmic resistivity (solid curve), Hall effect (dashed curve), and ambipolar diffusion (dotted curve) are also shown as functions of height at $x=0.3$\,au, $y=0$ (right panel). The long-dashed and dot-dashed curves show the coupling thresholds for toroidal and MRI fields, $\Omega H^2$ and $v_{a,z}^2\Omega^{-1}$, respectively. }
\end{figure*}

Fig. \ref{fig:diffusivity} shows diffusivity profiles for the gap. The left and centre panels show 2D slices through the gap, with the colour scale showing the field perpendicular diffusivity $\eta_{\perp}=(\eta_P^2+\eta_H^2)^{1/2}$, along with logarithmically spaced contour levels. The left panel shows the top-down view at the midplane, and the centre panel shows the edge-on profile  at $y=0$. The right panel shows the vertical profiles of the transport coefficients of Ohmic resistivity, the Hall effect and ambipolar diffusion, along with the coupling threshold for  toroidal and MRI fields.

Ohmic resistivity traces the high density structures such as the spiral arms and circumplanetary disc and is lowest in the evacuated regions to the upper-right and lower-left of the circumplanetary disc. It decreases with height in keeping with the exponential density profile. Hall drift is relatively constant with height above the midplane and dominates up to $z=2.5\,H_p$. Ambipolar diffusion is strongest in the low density atmosphere where neutrals are too sparse to influence electron and ion motion. 

Non-ideal effects are orders of magnitude below the level at which  the toroidal field decouples, except in the circumplanetary disc. Here the field is uncoupled to the midplane gas flow, but is anchored to, and transported by,  gas in the overlying coupled region. 

Non-ideal effects are below the MRI suppression threshold in the disc atmosphere, $z\gtrsim H_p$. The turbulent region is limited by the weak field condition [equation (\ref{eq:betaz_constraint})], rather than diffusion. Magnetic pressure is weaker than gas pressure at $z\lesssim2H_p$, and the field is weak enough to bend. The strong field region reaches towards the midplane near the planet where  field enhancement from flux-conservation is greatest.

\subsection{Magnetic field strength and geometry}
\label{subsec:magnetic_field_results}

\begin{figure*}
\begin{center} 
\includegraphics[width=0.60\textwidth]{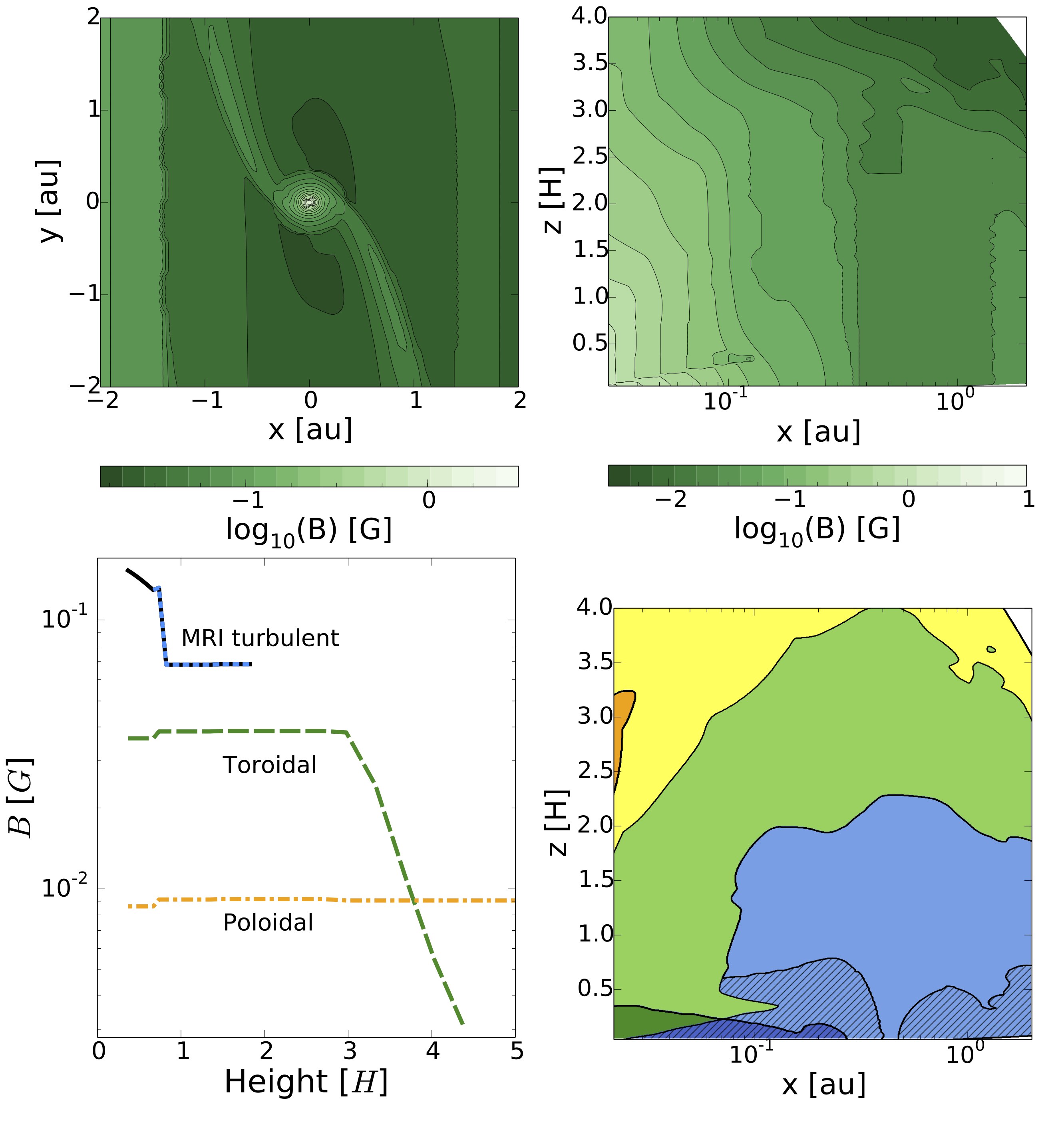}
\end{center}
\caption{\label{fig:magnetic_field} Properties of the estimated magnetic field in the gap for the standard parameter set $f_{dg}=10^{-4}$, $f_{\Sigma}=1$, $f_B=1$. Top-left and -right panels show contour plots of the field strength, $B$, at the midplane and $y=0$, respectively. Bottom-left panel shows the vertical profile of  field components at  $x=0.3\,$au, $y=0$. The vertical, toroidal,  MRI field with $s=1$, and MRI field with $s=-1$ are shown as the dot-dashed, dashed, solid, and dotted curves respectively.  Bottom-right panel shows the edge-on view of the field geometry, colour-coded according to the dominant field component for $s=1$: (a) poloidal field (yellow), (b)  toroidal field (green), and (c) MRI turbulent field (blue). Orange, dark green, and dark blue regions indicate where the Hall effect sustains and influences the toroidal field. Hatched MRI unstable regions show where Hall effect dominates, so that the field is predominantly turbulent if $s=1$, and toroidal if $s=-1$.}
\end{figure*}

Fig. \ref{fig:magnetic_field} shows the magnetic field strength and geometry.  The top panels are contour plots, in the top-down (top-left panel) and edge-on (top-right panel) views, with colour scale showing $\log_{10}(B)$,  calculated for $s=1$. The bottom-left panel shows vertical profiles for the poloidal, toroidal and turbulent magnetic field components evaluated at $x=0.3\,$au$, y=0$. A turbulent field is only present in the active zone between $0.5\lesssim z\lesssim\,2\,H_p$ for $s=-1$ and  for $z\lesssim\,2\,H_p$ for $s=1$.  The bottom-right panel is an edge-on view of the gap, colour coded according to the dominant field component. The gap is divided into the following regions: (a) poloidal (yellow), (b) toroidal  (green), and (c) MRI turbulent (blue) field. Orange, dark green,  and dark blue regions indicate the corresponding regions where the Hall effect maintains coupling of the toroidal component, and so the field may be counter-wrapped when $B_z$ is aligned with the rotation axis. Hatched blue regions are Hall-MRI unstable and so they are predominantly turbulent if $s=1$,  or predominantly toroidal if $s=-1$. 

The field strength is relatively uniform in the gap but flux conservation enhances the field in the circumplanetary disc. A toroidal field would easily couple and permeate the gap, but is limited by magnetic pressure in the disc atmosphere, and so the field would be increasingly poloidal with height. The gap would be turbulent between $z\sim0.5$--$2\,H_p$, except in the circumplanetary disc where the field is too strong. The Hall effect extends the turbulent region to the protoplanetary disc and circumplanetary disc midplanes if $s=1$, however the midplanes are MRI stable if $s=-1$. The upper boundary of the turbulent zone is limited by the weak field requirement, rather than non-ideal effects.  

\begin{figure*}
\begin{center}
\includegraphics[width=0.9\textwidth]{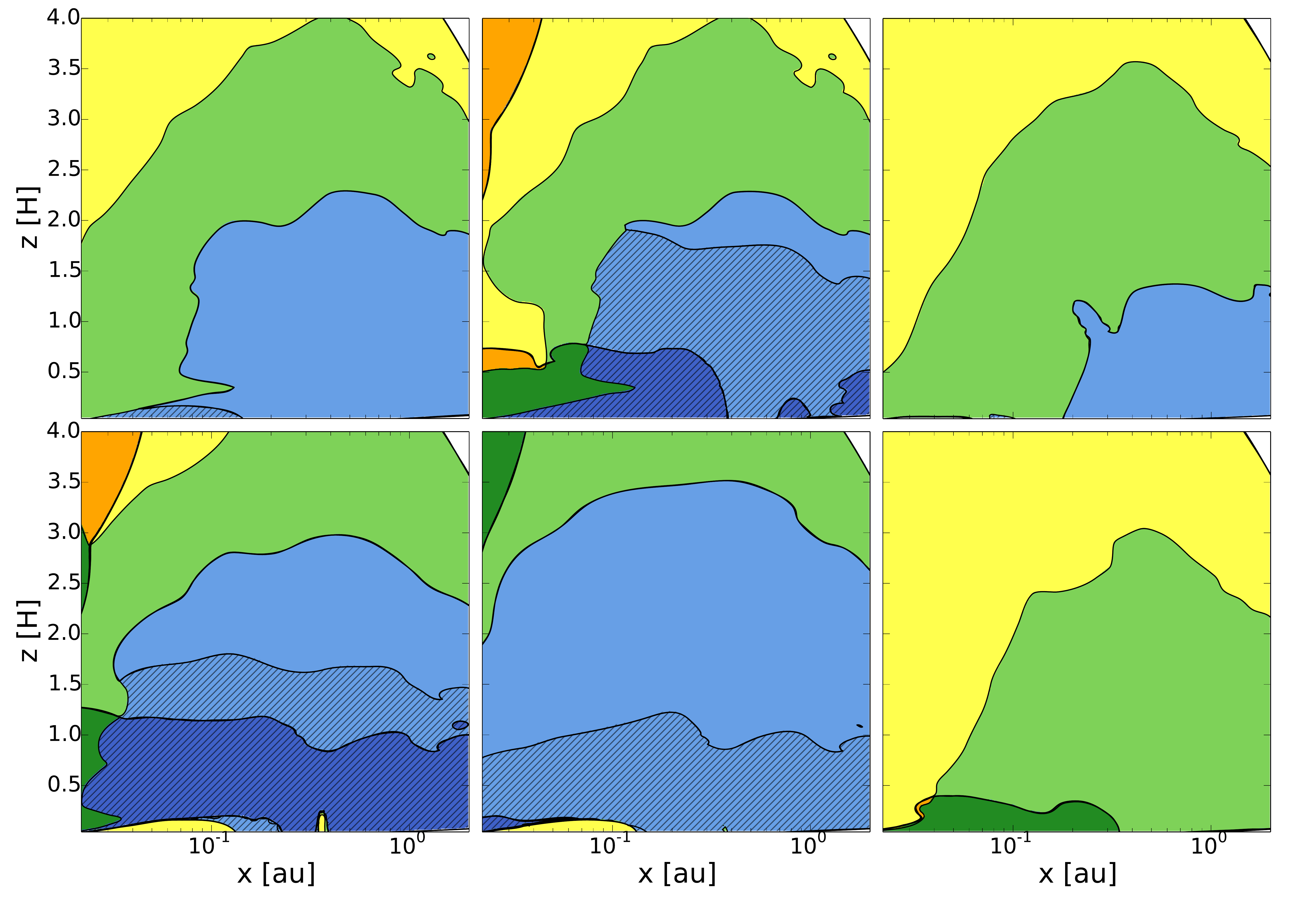}
\end{center}	
\caption{\label{fig:summary} Same as bottom-right panel in Fig. \ref{fig:magnetic_field}, except varying the dust-to-gas mass ratio $f_{dg}=0, 10^{-2}$ (top- and bottom-left panels), column density enhancement factor $f_{\Sigma}=0.1, 10$ (top-and bottom-centre panels), and magnetic field enhancement factor $f_{B}=0.1, 10$ (top- and bottom-right panels). }
\end{figure*}

Fig. \ref{fig:summary} shows the effect of varying parameters on the field geometry: $f_{dg}=0$, $10^{-2}$ (top- and bottom-left panels),  $f_{\Sigma}=0.1$, $10$ (top-and bottom-centre panels), and $f_{B}=0.1$, $10$ (top- and bottom-right panels). In general, we find that the gap is either predominantly toroidal, ideal MRI, or Hall-MRI unstable. 
 The upper MRI boundary  only depends on $\beta_z$ through the column  density and magnetic field strength, whereas the boundary of the Hall-MRI layer is controlled by the height at which the Hall drift exceeds the MRI threshold, $\Omega v_{a,z}^{-2}$. Thus, the Hall-MRI region extends by lowering the ionisation fraction, caused either through enhanced surface density attenuating ionising radiation (e.g., for $f_{\Sigma}>1$), or by grains soaking up free electrons (e.g., for $f_{dg}=10^{-2}$). Lowering the magnetic field strength reduces the Alfv\'{e}n speed, and consequently lowers the MRI threshold to below the Hall drift (e.g., for $f_{B}<1$). 
 
On the other hand, the circumplanetary disc may host a wide range of conditions.  Here the Hall-MRI zone in the circumplanetary disc is resilient to changes in $f_{dg}$, but  recedes if $\beta_z$ increases through enhanced density or reduce field strength, so that the field is predominantly toroidal. The field would be predominantly poloidal if  Ohmic resistivity exceeds the Hall effect, either by enhanced resistivity through higher density, or reduced Hall EMF from a lower field strength. However, further analysis into the potential for MRI turbulence in the circumplanetary disc will rely on models which are not affected by numerical viscosity.

\section{Implications for MHD gap modelling}
\label{subsec:magnetic_forces_results}
In this section we consider the implications of including non-ideal effects in MHD modelling of gas flow. We determine the extent to which diffusion limits small scale field gradients to calculate the minimum field bending radius. We discuss the implications of diffusion-limited field bending on the resolution needed for simulations. We also estimate the strength of large-scale magnetic forces, to compare with the hydrodynamical forces included in the TOM12 simulation. This allows us to determine where, if at all, magnetic forces have the most influence on gas flow.

\subsection{Minimum field-gradient length-scale} \label{subsec:field_tangling_length}
Non-ideal effects, particularly Ohmic resistivity and ambipolar diffusion, can wash out magnetic field gradients and limit the field bending length-scale, $L_B$. Here we calculate the smallest length-scale that the field can bend on, $L_{\text{min}}$, given the level of magnetic resistivity and diffusion throughout the gap as calculated in Section \ref{sec:coupling_ls_results}. The role of the Hall effect on magnetic field gradients is highly uncertain, being able to enhance or resist gradients. Therefore we neglect its effect in this simple calculation, but note the potential key impact it can have on the large-scale structure we consider here.

Just as equations (\ref{eq:MRI_criterion1}),  (\ref{eq:MRI_criterion2})--(\ref{eq:ideal_toroidal_criterion1}) specify the maximum diffusivity which permits the gradients needed for a turbulent or toroidal field, we can invert this relationship to specify the minimum bending length-scale given the magnetic diffusivity. This is applicable to large-scale, quasi-static magnetic field structures (as opposed to rapidly fluctuating turbulence), for which magnetic induction is balanced by magnetic diffusion. We derive the relation between magnetic diffusion and a general field geometry using the induction equation [equation (\ref{eq:induction})]. 
While it is generally necessary  to treat each component of the induction equation separately, we can estimate the diffusion limit for the minimum gradient length-scale, 
\begin{equation}
\eta_P<vL_B\left(\frac{L_B^{-1}+L_v^{-1}}{L_B^{-1}+L_{\eta}^{-1}}\right)\label{eq:general_eta_threshold}. 
\end{equation}
We calculate the velocity and diffusivity gradient length-scales, $L_v$, and $L_{\eta}$, using 
\begin{equation}
L_f=f/\lvert\nabla f \rvert,\label{eq:length_scale}
\end{equation}
with $f$ set to the relevant fluid velocity or Pedersen diffusivity, respectively.
Note that as we use the velocity in the frame orbiting with the planet to calculate $L_v$,  we remove the Keplerian component with respect to the star, which acts as a large constant offset in equation (\ref{eq:length_scale}). 

We invert equation (\ref{eq:general_eta_threshold}) to determine the \textit{minimum} field gradient length-scale $L_{\text{min}}$:
\begin{equation}
L_{\text{min}}=\frac{2\eta_P}{v}\left[1-\frac{\eta}{vL_{\eta}} +\sqrt{\left(1-\frac{\eta}{vL_{\eta}}\right)^2+\frac{4\eta}{vL_v}}\,\right]^{-1}.\label{eq:min_tangling_length}
\end{equation}

\begin{figure*}
\begin{center}
\includegraphics[width=0.98\textwidth]{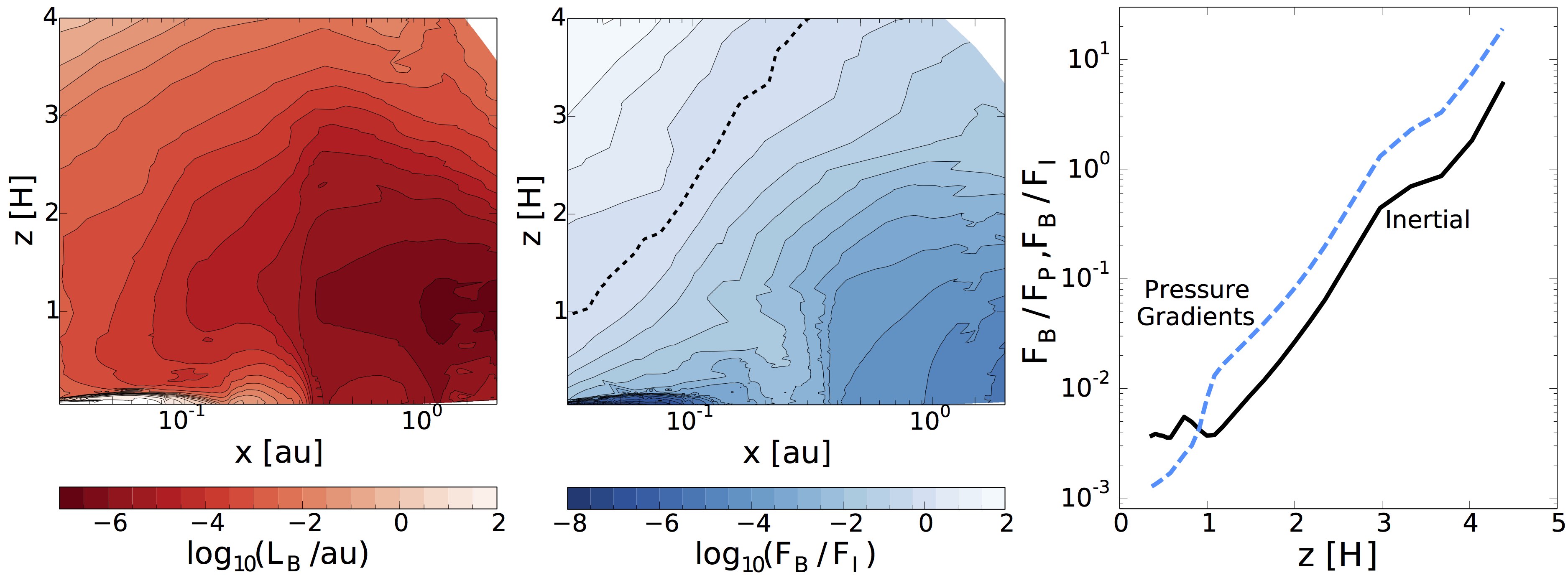} 
\end{center}
\caption{\label{fig:implications} Left panel shows contours of the minimum field bending-scale, $L_{\text{min}}$, shown in the edge-on view at $y=0$. Centre and right panels shows the strength of magnetic forces relative to hydrodynamical forces. The centre panel shows  the logarithm of the ratio of the magnetic force to the inertial force, $\log_{10}(F_B/F_I)$ at $y=0$. The  dashed contour shows the location of $F_B=F_I$. The right panel shows vertical profiles of $\log_{10}(F_B/F_I)$ and $\log_{10}(F_B/F_P)$, the logarithm of the ratio of the magnetic force to the inertial and pressure gradient forces at $x=0.3$\,au, $y=0$ as the solid and dashed curves, respectively. }
\end{figure*}

Figure \ref{fig:implications} shows a contour plot of  $\log_{10}(L_{\text{min}}/$au$)$ at $y=0$, with logarithmically spaced contour level. The minimum field bending length-scale follows $L_{\text{min}}\propto \eta_P/v$ everywhere except the circumplanetary disc where the $\eta_P$ is large and $L_v$, $L_{\eta}$ are small. It increases toward the planet, and is smallest at $z\sim H_p$ in the protoplanetary disc where the diffusivity is lowest. 

This can be used to gauge the minimum resolution required for MHD simulations. For example, as the minimum field bending length scale is  large in the circumplanetary disc, the resolution is determined by velocity and pressure gradients length-scales instead. In the protoplanetary disc, significantly higher resolution is needed to resolve turbulence which can be tangled on a very small length scales. The Hall effect can pose an additional challenge to simulations as it is dissipationless and so can introduce small-scale field structure (e.g. \citealp{2002ApJ...577..534S}, \citealp{2013MNRAS.434.2295K, 2014A&A...566A..56L, 2014MNRAS.441..571O, 2015ApJ...798...84B}).

\subsection{Relative strength of magnetic forces} \label{subsubsec:magnetic_forces}

Large-scale magnetic fields have the potential to influence  gas dynamics, rather than being merely passively drawn along with the gas. We touched on this in Section \ref{subsec:magnetic_field} when we compared the toroidal component of magnetic pressure  with gas pressure.  For example, circumplanetary jets have been launched in ideal and resistive MHD simulations \citep{2006ApJ...649L.129M, 2013ApJ...779...59G}.  

The TOM12 simulation is not magnetised; nevertheless we can estimate the size of magnetic forces and compare them with  hydrodynamic forces  in the simulation. This will indicate where,  if at all,  they have the greatest potential to alter the balance of forces and so influence the gas flow. It also acts as a consistency check, validating (or otherwise) our adoption of a hydrodynamic simulation as the underlying model.

The full momentum equation, including both hydrodynamic and magnetic forces is
\begin{eqnarray}
\frac{1}{\rho}\frac{\partial\bold{v}}{\partial t}+\left(\bold{v}\cdot\nabla\right)\bold{v} &=&- 2\Omega_p \bold{\hat{z}}\times \bold{v}-\Omega_p^2\bold{\hat{z}}\times\left(\bold{\hat{z}}\times\bold{r}\right)-\nabla \Phi\nonumber\\
&& -\frac{1}{\rho}\nabla p+\frac{1}{4\pi\rho}\left(\nabla  \times \bold{B}\right)\times \bold{B},\label{eq:mom_eqn}
\end{eqnarray}
where $\bold{r}=(x,y,z)$ and $\Phi$ is the combined gravitational potential from the star and planet:
\begin{equation}
\Phi=-\frac{G M_*}{\lvert\bold{r}-\bold{r_*}\rvert}-\frac{G M_p}{\lvert\bold{r}\rvert}.
\end{equation}
The hydrodynamic terms in equation (\ref{eq:mom_eqn}) are the time derivative and inertial force on the left hand side, and the Coriolis, centrifugal force, gravitational, and pressure gradient forces on the right hand side. These are calculated in the frame co-orbiting with the protoplanet in the TOM12 simulation. 
We neglect the time derivative because it is small with the simulation in almost steady state. 

The addition of the magnetic force [the final term on the right hand side of equation (\ref{eq:mom_eqn})] will inevitably change the balance of forces, however the impact will be negligible unless the magnetic force is strong compared to the inertial force. We calculated the inertial force  $F_I=\lvert\left(\bold{v}\cdot\nabla\right)\bold{v}\rvert$ from the TOM12 simulation and its associated gradient length scale
\begin{equation}
L_I=\frac{\bold{v}^2}{\lvert\left(\bold{v}\cdot\nabla\right)\bold{v}\rvert},
\end{equation}
using second-order finite differencing, with good convergence compared with first-order differencing.

We estimate the magnetic force from large-scale non-turbulent fields using,
\begin{equation}
F_B\equiv\frac{1}{4\pi\rho}\lvert\left(\nabla  \times \bold{B}\right)\times \bold{B}\rvert\sim\frac{v_a^2}{L_B}.
\end{equation}
We do this by using the Alfv\'{e}n speed calculated for the total, non-turbulent field strength. Our magnetic force estimate relies on a knowledge of the field gradient length-scale, $L_B$. We expect that, in general, $L_B$ traces the fluid gradient length-scales because either the field is dragged along with the fluid or visa versa. If this is the case, we take $L_B=L_I$, and it  cancels from the ratio of the magnetic and hydrodynamic forces $F_B/F_I$. If, instead, the gas varies over a shorter length-scale than diffusion permits the  field to bend,  we take $L_B=L_{\text{min}}$ as calculated in Section \ref{subsec:field_tangling_length}. In this way, we compare the magnetic force over the same scale as the hydrodynamic forces, to the extent that diffusion permits.   We have not included the the force arising from effective MRI turbulent viscosity, as it is already the subject of extensive study within the literature (see \citealp{2014prpl.conf..411T}). 
 
We also consider the effect magnetic forces have on the scale height by comparing them with the pressure gradient force, $F_P=\lvert(\nabla p)/\rho\rvert$, with length-scale $L_P$ calculated using equation (\ref{eq:length_scale}).

The centre-panel of Fig. \ref{fig:implications} shows a contour plot of the logarithm of the ratio of the magnetic force to the inertial force, $\log_{10}(F_B/F_I)$ at $y=0$. The dashed contour denotes $F_I=F_B$.
The right panel shows the vertical profile of the ratio of the magnetic force to the inertial,  and pressure gradient forces. Magnetic forces are strongest in the disc atmosphere, exceeding the inertial and pressure gradient forces above $z\gtrsim\,2.5$--$3.5\,H_p$, where they are able to influence the protoplanetary disc scale-height.

Magnetic forces are weakest in the circumplanetary disc where strong magnetic diffusion limits the field bending length scale to the diffusion length scale. Although  artificial diffusion lessens the reliability of the circumplanetary disc structure in the TOM12 simulation, KW14 shows that magnetic  diffusion is strong across a range of circumplanetary disc models, owing to the column high density shielding the disc from external ionising radiation. The strong diffusion increases the minimum field bending length scale, thereby reducing the magnetic force. This indicates that large-scale magnetic forces cannot yield any significant accretion in the circumplanetary disc, which must come from the MRI or Gravitational Instability-MRI limit-cycles (e.g., \citealp{2011ApJ...740L...6M, 2012ApJ...749L..37L}). 

\section{Discussion}
\label{sec:discussion}

In this paper we examined the importance of non-ideal effects in determining the magnetic field structure in a gap surrounding a giant protoplanet. We modelled the gap using a snapshot from the pure-hydrodynamical gap-crossing simulation by \citet{2012ApJ...747...47T}. Our approach was to use this snapshot as a basis to a posteriori estimate key MHD quantities semi-analytically, which would otherwise be very challenging to incorporate into simulations.  We calculated the  ionisation fraction produced by cosmic-rays,  stellar X-rays and radioactive decay, including the effect of grains. We calculated Ohmic resistivity, ambipolar diffusion and Hall drift to determine whether an MRI field could be generated and if a toroidal field could couple to the gas flow. We estimated the magnetic field strength in the protoplanetary disc from inferred accretion rates, and determined the gap field strength from flux-freezing. 

We found that a magnetic field would be easily drawn from the protoplanetary disc into the gap and circumplanetary disc. A toroidal field permeates the gap, but is weakened by expansion from magnetic pressure above $z>2$--3\,$H_p$. The gap is MRI unstable at $z>0.5\,H_p$ with turbulence extending down to the midplane if the vertical component of the magnetic field is along the rotation axis (i.e., if $s=1$). If however, the vertical field and rotation axis are anti-aligned, the conditionally unstable regions are non-turbulent. As protoplanetary discs exhibit a range of field/rotation axis misalignment angles we expect the size of the MRI turbulent region to differ between systems, with significant implications for the flow dynamics \citep{2013ApJ...768..159H, 2013ApJ...767L..11K}. 

This direction dependence of the MRI originates with the Hall effect. The Hall effect is strong below $z<2\,H_p$ and  plays an important role in generating MRI instability. We  found that it can also facilitate coupling of a toroidal field, despite strong magnetic Pedersen diffusion, and influences the orientation of the toroidal component. This may lead to counter-wrapping of the toroidal field if $s=1$, where the field drift opposes the Keplerian gas flow. Further simulations are needed to probe the role of the Hall effect in this system. 

By testing the sensitivity of the calculations to  dust content, column density, and magnetic field strength, we found that the gap was generally susceptible to the Hall- or ideal-MRI. On the other hand, the MRI in the circumplanetary disc  may be quite sensitive to disc conditions and could have a turbulent, toroidal or vertical field. As the circumplanetary disc evolves, the disc may experience a range of different field configurations in keeping with the varying column density and dust content. Understanding the evolution of a circumplanetary disc is key for satellite formation studies, which is believed to be the formation site of moons. Bimodality in circumplanetary disc dynamics caused by the Hall effect may transfer to the growth of  moons within the system. For example, turbulent heating will effect the location of the circumplanetary ice-line. 

Finally, we have calculated the minimum magnetic field gradient length-scale limited by magnetic diffusion. Magnetic diffusion resists field gradients, and so minimal field bending is permitted in the circumplanetary disc where Ohmic resistivity is strong. We find that large-scale (non-turbulent) magnetic forces are unable to drive accretion in circumplanetary discs. Turbulence aside, we found the large scale-flow features are well modelled by a hydrodynamical fluid as large scale magnetic forces as small outside the protoplanetary disc atmosphere. Here, magnetic coupling is strong and so a strong field may be able to produce a jet, winds, or other variability. 

In summary, we find that the magnetic field surrounding a giant protoplanet is mostly toroidal, with large bands of ideal- and Hall-MRI turbulent zones. The magnetic field geometry is dependent on the orientation of the vertical field component, established during the collapse of the protostellar core. Non-ideal effects are important in the gap and need to be included in future MHD simulations of gap crossing.

\section*{Acknowledgements}
We are grateful to Takayuki Tanigawa for the generous provision of the simulation data, along with assistance in using and interpreting the simulation. We also thank Pandey, and  Oliver Gressel for valuable discussions, and Philippa Browning, Michael Keith, and the anonymous referee for helpful comments on the manuscript. The authors thank the Niels Bohr International Academy for hospitality. This work was supported in part by the Australian Research Council grant DP130104873. S.K. further acknowledges the support of an Australian Postgraduate Award, funding from the Macquarie University Postgraduate Research Fund scheme and the International Space Science Institute. This research has made use of NASA's Astrophysical Data System.

\bibliographystyle{mn2e}
\bibliography{references}

\appendix
\section{Weak-field condition for Hall-MRI}\label{appendix:weak_field}
Developing and sustaining MRI turbulence requires two conditions: (i) firstly, that field perturbations must grow fast enough, and (ii) secondly, the wavelength of the fastest-growing turbulent mode $\lambda$, must be contained within a disc scale-height (i.e., $\lambda\le H$). Here we develop the weak-field condition for Hall-MRI. The wavelength of the fastest-growing Hall-MRI mode is [see equation (B14) of \citealp{2012MNRAS.422.2737W}]
\begin{equation}
\lambda=\frac{\pi}{\nu}\left[3s\eta_H\Omega-4\eta_P\nu-4v_{a,z}^2+\frac{10v_{a,z}^2\Omega^{2}}{\nu^2+\Omega^{2}}\right]^{\frac{1}{2}},
\label{eq:weak_field_hall}
\end{equation}
where $\nu$ is the growth rate of the fastest-growing mode. In the Hall MHD limit, $\eta_P=0$, the maximum growth rate attains the ideal rate $\nu=\frac{3}{4}\Omega^{-1}$ \citep{2012MNRAS.422.2737W}. Using this result  in strong-Hall limit, $\eta_H\Omega v_{a,z}^{-2}\gg1$, the wavelength of the fastest-growing mode is approximately 
\begin{equation}
\lambda\approx2\pi\sqrt{\frac{\eta_H}{\Omega}},
\end{equation}
up to a constant factor of order unity. 

 Ensuring that the two MRI conditions [(i), (ii) above] are met allows us to bracket the Hall-drift:
\begin{equation}
\frac{v_{a,z}^2}{\Omega}\le\eta_H\le\Omega\left(\frac{H}{2\pi}\right)^,
\end{equation}
 which, with the thin-disc approximation $H=c_s/\Omega$, leads to the weak-field limit
\begin{equation}
\beta_z=\frac{2c_s^2}{v_{a,z}^2}\ge8\pi^2.
\end{equation}
This is identical to that in the ideal and resistive MRI regimes [equation (\ref{eq:betaz_constraint}); \citealp{2011ApJ...742...65O}].

\section{Non-ideal effects acting on  a toroidal field}
\label{appendix:toroidal_hall}
 Here we determine the coupling threshold for a toroidal field using the azimuthal component of the induction equation. We consider a Keplerian disc with velocity, $\bold{v}=\Omega r\boldsymbol{\hat{\phi}}$, independent of height, and uniform transport coefficients $\eta_O, \eta_H, \eta_A$ for simplicity. 

Shear generates a toroidal component from a  poloidal field in the inductive term: $\left[\nabla\times(\mathbf{v}\times\mathbf{B})\right]_{\phi}=-\frac{3}{2}\Omega B_r$. The resistive term, $[-\nabla\times(\eta_O\nabla\times B)]_{\phi}=\eta_O(\nabla^2B_{\phi}-B_{\phi}/r^2)\sim\eta_O B_{\phi}/H^2$, is dominated by vertical gradients (see Section \ref{subsec:magnetic_field}; \citealp{1993ApJ...410..218W}), as are the Hall and ambipolar terms, which scale similarly. Comparing the two terms yields the coupling criterion\footnote{This threshold is a factor of $10\left(H/r\right)^{-2}\sim5\times10^3$ lower than \citet{2008ApJ...679L.131T}, as we allow for vertical gradients in the non-ideal terms.}
\begin{equation}
\eta<\Omega H^2, \label{eq:coupling_criterion_appendix}
\end{equation} 
where $\eta=\eta_P , \eta_H$.

How will the field evolve once it decouples from the gas motion? Recasting the induction equation to show the field line drift, $\bold{V_B}$, makes this clear \citep{2012MNRAS.422.2737W}:
\begin{equation}
 \frac{\partial \bold{B}}{\partial t}=\nabla\times\left\{\left(\bold{v}+\mathbf{V_B}\right)\times \bold{B}-\eta_O\left[(\nabla\times \bold{B})\cdot\hat{\bold{B}}\right]\hat{\bold{B}}\right\},
\end{equation}
where
\begin{equation}
\mathbf{V_B}=\frac{\eta_P}{B}\left(\nabla\times\bold{B}\right)_{\perp}\!\times \hat{\bold{B}}-\frac{\eta_H}{B}\left(\nabla\times\bold{B}\right)_{\perp}
\end{equation}
The azimuthal component of the drift velocity, $V_{B,\phi}$, controls the wrapping of the field lines with the shear. Neglecting terms of $\mathcal{O}(H/r)$, the azimuthal drift velocity is \citep{2012MNRAS.427.3188B}:
\begin{eqnarray}
V_{B,\phi}&=&\frac{\eta_P}{H}\frac{B_zB_{\phi,s}}{B^2}-\frac{\eta_H}{H}\frac{B_{r,s}}{B}\\
&\equiv&V_P+V_H,
\end{eqnarray}
where we have separated the Pedersen and Hall drift components. Pedersen drift always resists winding, tending to reduce $\lvert B_{\phi,s}\rvert$ and straighten field lines. On the other hand, the Hall component will enhance or resist winding according the sign of $B_{r,s}$. 

In the natural configuration for $s=1$ (i.e., $B_{r,s}>0$, and $B_{\phi,s}<0$ for $z>0$), $V_H<0$ and $V_P<0$ so that  the components cooperate in unwrapping the field. For dominant Hall, the unwrapping will overshoot, winding the field in the opposite direction  so that $B_{\phi,s}>0$.  For the anti-aligned field, $s=-1$, the natural configuration is  $B_{r,s}<0$, and $B_{\phi,s}>0$ above the midplane. Once again, Pedersen resists wrapping of the field lines and if it dominates the field will tend towards vertical. On the other hand, now Hall drift tends to enhance field motion along $\hat{v}_{\phi}$ so that even if  the field is somehow wound against the flow (i.e., $B_{\phi,s}>0$) both components cooperate to restore $B_{\phi,s}\ge0$ in equilibrium.

\begin{figure}
\begin{center}
\includegraphics[width=0.42\textwidth]{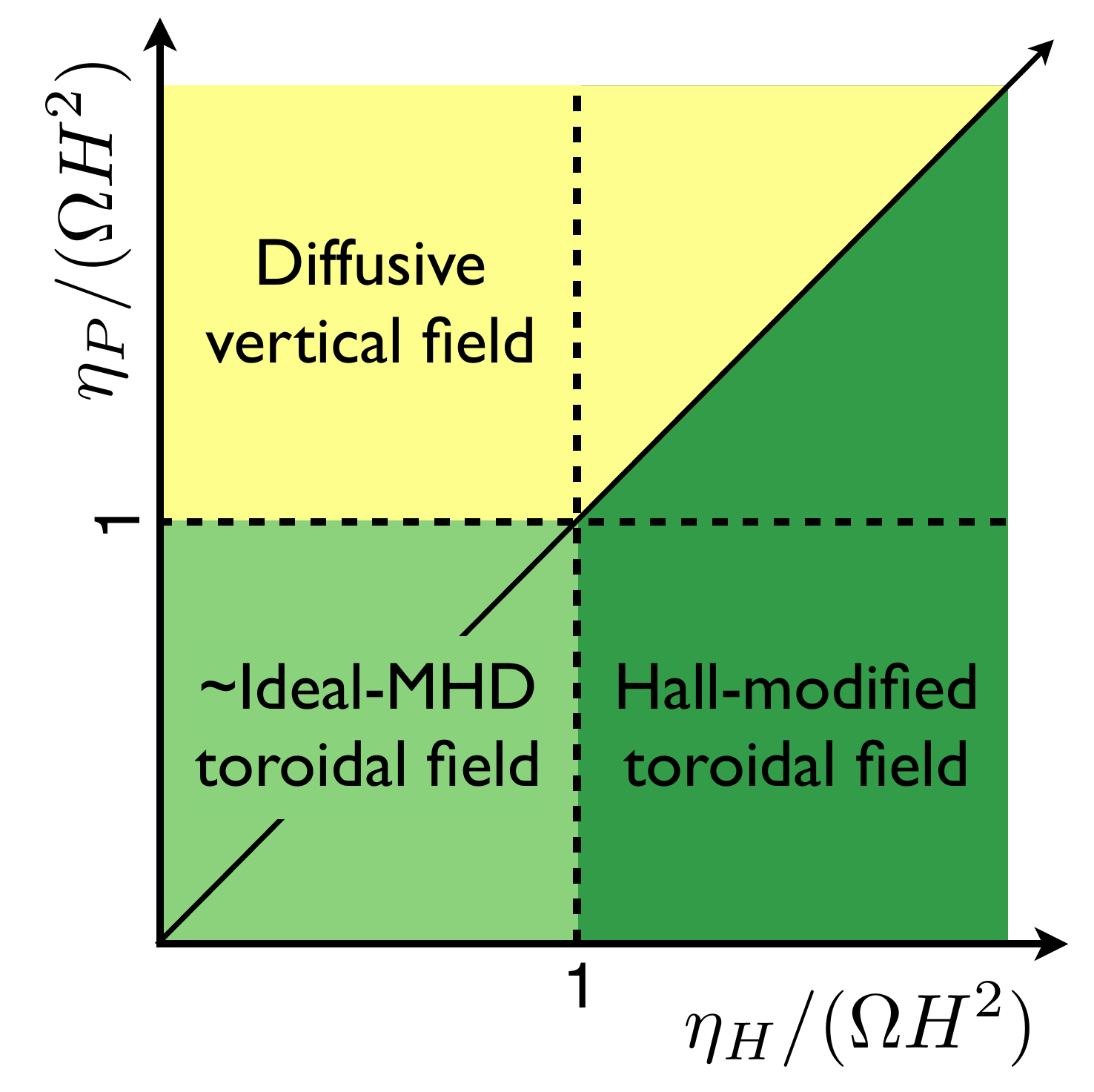} 
\end{center}
\caption{\label{fig:non_ideal_toroidal_summary} Regions of $\eta_H$--$\eta_P$ parameter space, shaded according to the toroidal field drift. Ideal MHD (light green) and Hall dominated MHD (dark green) will sustain a toroidal field, although the orientation of the field varies with the vertical field orientation in the Hall-MHD region. Pedersen diffusion only permits a vertical field in the yellow region.}
\end{figure}

This behaviour is summarised in Fig. \ref{fig:non_ideal_toroidal_summary}, which shows how the resulting field geometry varies in the $\eta_H$--$\eta_P$ parameter space. Non-ideal effects are weak in the lower-left quadrant (light green), with the toroidal field wound up by the disc so that $B_zB_{\phi}<0$. Pedersen diffusion dominates above the diagonal (yellow region), so that the field tends toward vertical. The Hall effect counteracts diffusion below the diagonal (dark green), so that the toroidal component is enhanced  for $s=-1$ or counter-wrapped when $s=1$.

\bsp
\label{lastpage}

\end{document}